\documentclass[journal=ancac3,manuscript=article]{achemso}
\usepackage{graphicx}
\usepackage{braket}
\usepackage{epsfig}
\usepackage{color}

\usepackage{multirow} 
\usepackage{float}    

\usepackage{amsmath} 
\usepackage{amssymb} 
\usepackage{ulem} 
\usepackage[dvipsnames]{xcolor} 

\usepackage{bm}

\newcommand{\onlinecite}[1]{\hspace{-1 ex} \nocite{#1}\citenum{#1}} 

\title{Single and double electron spin-flip Raman scattering in CdSe colloidal nanoplatelets}

\author{Dennis Kudlacik}
\affiliation{Experimentelle Physik 2, Technische Universit\"{a}t Dortmund, 44227 Dortmund, Germany}

\author{Victor F. Sapega}
\affiliation{Ioffe  Institute, Russian Academy of Sciences, 194021 St. Petersburg, Russia}

\author{Dmitri R. Yakovlev}
\affiliation{Experimentelle Physik 2, Technische Universit\"{a}t
Dortmund, 44227 Dortmund, Germany}
\alsoaffiliation{Ioffe Institute, Russian Academy of Sciences, 194021 St. Petersburg,
Russia}
\email{dmitri.yakovlev@tu-dortmund.de}

\author{Ina V. Kalitukha}
\affiliation{Ioffe  Institute, Russian Academy of Sciences, 194021 St. Petersburg, Russia}

\author{Elena V. Shornikova}
\affiliation{Experimentelle Physik 2, Technische Universit\"{a}t Dortmund, 44227 Dortmund, Germany}

\author{Anna V. Rodina}
\affiliation{Ioffe  Institute, Russian Academy of Sciences, 194021 St. Petersburg, Russia}

\author{Eugeniius L. Ivchenko}
\affiliation{Ioffe  Institute, Russian Academy of Sciences, 194021 St. Petersburg, Russia}

\author{Grigorii S. Dimitriev}
\affiliation{Ioffe  Institute, Russian Academy of Sciences, 194021
St. Petersburg, Russia}

\author{Michel Nasilowski}
\affiliation{Laboratoire de Physique et d'Etude des Mat\ifmmode
\acute{e}\else \'{e}\fi{}riaux, ESPCI, CNRS, 75231 Paris, France}

\author{Benoit Dubertret}
\affiliation{Laboratoire de Physique et d'Etude des Mat\ifmmode
\acute{e}\else \'{e}\fi{}riaux, ESPCI, CNRS, 75231 Paris, France}

\author{Manfred Bayer}
\affiliation{Experimentelle Physik 2, Technische Universit\"{a}t Dortmund, 44227 Dortmund, Germany}
\alsoaffiliation{Ioffe Institute, Russian Academy of Sciences, 194021 St. Petersburg,
Russia}

\keywords{Nanoplatelet, CdSe, excitons, spin-flip, Raman scattering}

\begin{document}

\begin{abstract}
CdSe colloidal nanoplatelets are studied by spin-flip Raman scattering in magnetic fields up to 5~T. We find pronounced Raman lines shifted from the excitation laser energy by an electron Zeeman splitting. Their polarization selection rules correspond to those expected for scattering mediated by excitons interacting with resident electrons. Surprisingly, Raman signals shifted by twice the electron Zeeman splitting are also observed. The theoretical analysis and experimental dependencies show that the mechanism responsible for the double flip involves two resident electrons interacting with a photoexcited exciton. Effects related to various orientations of the nanoplatelets in the ensemble and different orientations of the magnetic field are analyzed.   
\end{abstract}

\section{Introduction}

Colloidal semiconductor nanocrystals are of interest for various fields of chemistry, physics, biology, and medicine and are successfully used in various optoelectronic devices. Being synthesized from many different semiconductor materials, they can have different geometries, resulting in zero-dimensional quantum dots, one-dimensional nanorods or two-dimensional nanoplatelets (NPLs). CdSe NPLs demonstrated small inhomogeneous broadening~\cite{Ithurria2008}, as confirmed later also for NPLs of different composition.\cite{Nasilowski2016,Berends2017,Ithurria2012} Semiconductor NPLs can be considered as model systems to study physics in two-dimensions, similar to quantum well heterostructures and layered materials, like graphene or transition metal dichalcogenides.    

The spin physics of colloidal nanocrystals is still in its infancy compared to the rather mature field of spintronics based on epitaxially-grown semiconductor quantum wells and quantum dots. The properties of colloidal and epitaxial nanostructures can differ considerably due to much stronger confinement of charge carriers in colloidal structures leading to different properties such as the strongly enhanced Coulomb interaction (enhanced exciton binding energy and fine structure energy splitting), the possibility of photocharging and surface magnetism, the reduction of the phonon bath influence, etc. Experimental techniques widely used in spin physics of heterostructures can be, however, readily applied to colloidal quantum dots and NPLs. Among them are polarized photoluminescence in magnetic field including time-resolved spin dynamics~\cite{Liu2013,Yakovlev2018}, transient grating spectroscopy~\cite{Huxter2010}, magnetic circular dichroism~\cite{Gao2018,Fainblat2014,Muckel2018}, single dot spectroscopy,~\cite{Fainblat2016} and pump-probe time-resolved Faraday rotation.\cite{Gupta1999,Gupta2002,Fumani2013,Feng2017,Shornikova2018nl}  

The spin-flip Raman scattering (SFRS) technique is another valuable tool to address the spin structure and measure the $g$-factors of charge carriers and excitons. It has been used for semiconductors of different dimensionality: bulk,\cite{ToHo1968,Scott1972} quantum wells~\cite{SaCa1992,Sirenko1997,Debus2013} and quantum dots\cite{Debus2014,Debus2014b}. This technique has been also applied to CdS quantum dots in glass matrix~\cite{Sirenko1998} and to negatively charged CdSe/CdS  colloidal NPLs with thick shells.~\cite{Shornikova2018nl}   
In SFRS experiments the photon energy of the exciting laser is tuned to the exciton resonance, which serves as an intermediate state for the light scattering. As the experiments are performed in an external magnetic field, the spin-flip of a charge carrier (in most cases it is an electron) or of an exciton is accompanied by its transition between the Zeeman-split sublevels. Accordingly, the Raman scattered light has components shifted from the laser photon energy to the Stokes or anti-Stokes spectral regions by the amount of the Zeeman splitting. The Raman shift corresponding to the Zeeman splitting can be used for evaluation of $g$-factors. Also, the polarization properties of the Raman signals provide new insights into the responsible flip mechanisms. In low-dimensional structures the SFRS intensity is drastically increased by the presence of resident quantum-confined electrons that interact with photogenerated excitons.~\cite{Debus2013,Debus2014,Debus2014b}

In this paper we use the spin-flip Raman scattering technique to study the spin properties of CdSe colloidal nanoplatelets. By measuring ensembles of NPLs having various orientations in the Faraday and Voigt geometries of the external magnetic field and analyzing the polarization properties of the Raman spectra we determine the electron $g$-factor and its anisotropy. We develop a theory of Raman light scattering mediated by excitons interacting either with one or two resident electrons localized in a NPL. The theory explains well our experimental results and confirms that the unusual double-electron spin-flip found experimentally is provided by NPLs containing more than one resident electron.   

\section{Experimental results}

\begin{figure}[h!]
	\centering
	\includegraphics[width=15cm]{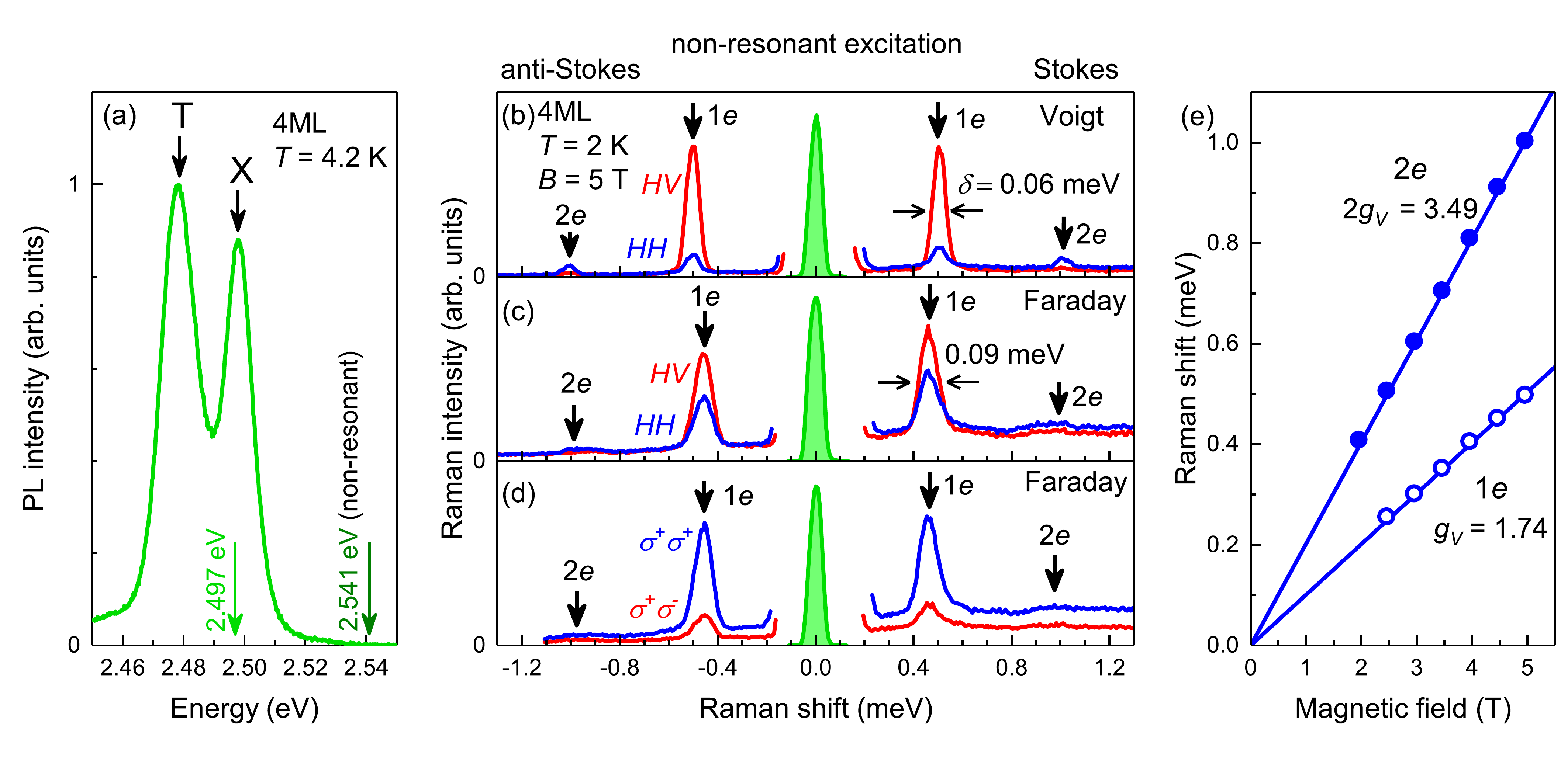}
	\caption{Photoluminescence and SFRS spectra of 4ML CdSe NPLs. 
	(a) PL spectrum measured for excitation with 3.061~eV photon energy at $T=4.2$~K. X and T lines correspond to emission of neutral excitons and negatively charged excitons (trions), respectively. Colored arrow indicates laser photon energy used for SFRS.
	(b,c,d) SFRS spectra under non-resonant excitation at 2.541~eV, power density $P=26$~Wcm$^{-2}$, magnetic field $B=5$~T and temperature $T=2$~K. Laser is shown by green line at zero Raman shift. (b) Voigt-geometry spectra measured in co- (blue) and cross- (red) linear polarizations. In Faraday geometry the spectra are measured in co- (blue) and cross- (red) linear polarizations [panel (c)] and in co- (blue) and cross- (red) circular polarizations [panel (d)].  
	(e) Magnetic field dependence of the Raman shift of the single, $\Delta^{1e}$ (open circles), and double, $\Delta^{2e}$ (solid circles), electron spin-flips measured in the Voigt geometry. Solid lines represent linear fits to the data with the respective $g$-factors.
   	}
	\label{SFRS_nr}
\end{figure}

In this paper we study three ensembles of CdSe NPLs with thicknesses of 3ML, 4ML and 5ML, their optical properties can be found in Ref.~\onlinecite{Shornikova2018ns}. For in-depth demonstration of the SFRS technique we choose 4ML sample. Its photoluminescence (PL) spectrum shown in Figure~\ref{SFRS_nr}(a) has two lines. We attribute the high-energy line X to radiative recombination of neutral excitons and the low-energy line T to negatively charged excitons (trions). The line identification is made on the base of spectral shifts in emission and absorption spectra, characteristic recombination dynamics and their dependence on temperature and magnetic field, polarization properties of emission in high magnetic fields.\cite{Shornikova2018ns,Shornikova2018nl,Shornikova2019nn}.

SFRS spectra of the 4ML NPLs are shown in Figure~\ref{SFRS_nr}(b-d). They are measured in magnetic field of $B=5$~T, applied in the Voigt and Faraday geometries. The laser energy is set to 2.541~eV in the high energy tail of the exciton PL band, i.e. corresponding to non-resonant excitation. The spectra are measured in the vicinity of the laser line (green line) towards the lower energy (Stokes, positive Raman shift) and higher energy (anti-Stokes, negative Raman shift) sides. They are composed of a broad background by photoluminescence and the narrow lines of the SFRS signals: One can see two Stokes and two anti-Stokes lines. We assign the lines with the Raman shift of 0.5~meV (Voigt geometry, Figure~\ref{SFRS_nr}(b)) to the single-electron SFRS and thus label them as 1$e$. The lines with the twice larger shift of 1~meV are consistently related to the SFRS process involving the spin-flips of two electrons, and are accordingly labeled as 2$e$.  This assignment is confirmed by the linear shifts of these lines with increasing magnetic field, Figure~\ref{SFRS_nr}(e). Linear fits by the forms $\Delta^{1e}=g_{\rm V}\mu_B B$ and $\Delta^{2e}=2g_{\rm V} \mu_B B$, where $\mu_B$ is the Bohr magneton, give $g_{\rm V}=1.74$ for the Voigt geometry. Importantly, their extrapolation to zero field reveals a negligible Raman shift offset. It should be noted that SFRS shift can have nonzero offset, which is an evidence of exchange interaction of the involved particles, e.g., the bright-dark exchange splitting of the exciton states.\cite{Debus2014} In the Faraday geometry, the similar magnetic field shifts of the $1e$ and $2e$ lines are characterized by a smaller $g$ factor $g_{\rm F}=1.59$. The width of 1$e$ SFRS line is dependent on the magnetic field direction: $\delta=0.06$~meV in the Voigt configuration and 0.09~meV in the Faraday configuration, Figure~\ref{SFRS_nr}(b,c). Note also that the intensity of the $2e$ lines is by an order of magnitude smaller compared to the $1e$ lines, as can be expected for a higher-order process involving two electron spin-flips instead of one. The observation of the double-electron SFRS is both surprising and remarkable, a double spin-flip transition had been previously reported only for donor-bound electrons in bulk CdS crystals\cite{Scott1972}, but not in a low-dimensional system. 

Depending on the mutual orientations of the magnetic field $\textbf{B}$ and the light wave vector $\mathbf{k}$, the SFRS lines are strongly or weakly polarized. In the Voigt geometry ($\textbf{B} \perp \textbf{k}$), the $1e$ line is predominantly observed for cross linear polarizations ($HV$ or $VH$), Figure~\ref{SFRS_nr}(c) and Supporting Information, Figure S3. Hereafter, the notations $H$ and $V$ are used for the horizontal and vertical orientation of the photon electrical vector ${\mathbf E}$ with respect to the magnetic field direction. Note that in our experiment the co-polarizations $HH$ and $VV$ are equivalent, the same is valid for the cross-polarizations $HV$ and $VH$. In contrast to the $1e$ line, the $2e$ line is more intense for parallel polarizations ($HH$ or $VV$). The ratio of intensities of the $1e$ and $2e$ lines measured in $HV$ and $HH$ polarizations amounts to $I^{1e}_{HV}/I^{1e}_{HH}=7$ and $I^{2e}_{HV}/I^{2e}_{HH}=0.3$. In Faraday geometry, the $1e$ line is predominantly circularly co-polarized $I^{1e}_{\sigma^+\sigma^+}/I^{1e}_{\sigma^+\sigma^-}=5$, while the $2e$ line has almost no circular polarization, Figure~\ref{SFRS_nr}(d). 

\begin{figure}[h!]
	\centering
	\includegraphics[width=15cm]{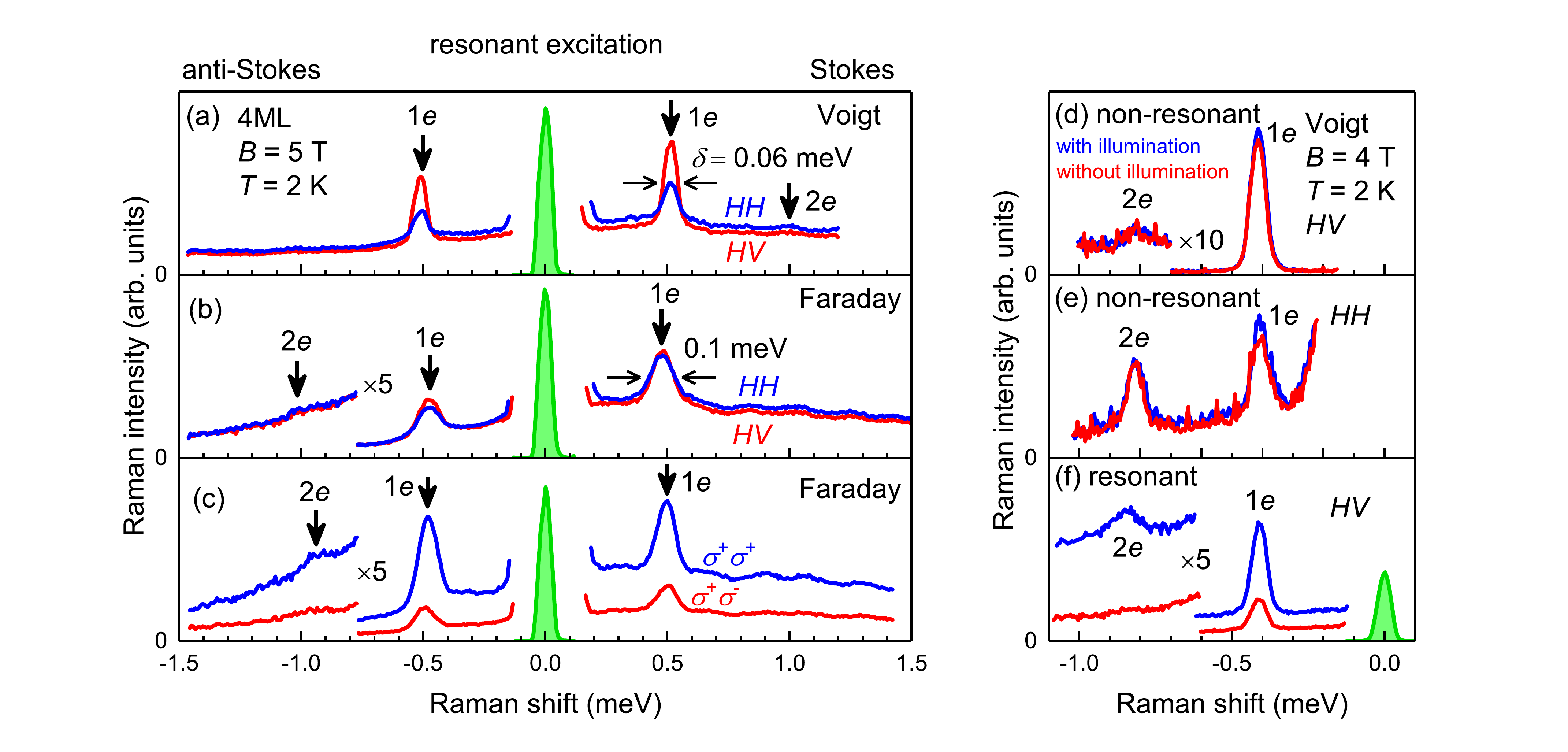}
	\caption{(a,b,c) SFRS spectra of 4ML CdSe NPLs measured under resonant excitation at 2.497~eV  (see Figure~\ref{SFRS_nr}(a)), power density $P=20$~Wcm$^{-2}$,  $B=5$~T and $T=2$~K. (a) Voigt-geometry spectra measured in co- (blue) and cross- (red) linear polarizations. Faraday-geometry spectra measured in co- (blue) and cross- (red) linear polarizations [panel (b)] and in co- (blue) and cross- (red) circular polarizations [panel (c)]. 
(d,e,f) SFRS spectra measured in cross-polarizations (d,f) and co-polarizations (e) for non-resonant 2.541~eV (d,e) and for resonant 2.497~eV (f) excitation. Blue and red spectra are measured with and without additional illumination, respectively. The illumination photon energy is 3.061~eV with a power density $P_{\rm ill}=1.3$~Wcm$^{-2}$. }
	\label{SFRS_res}
\end{figure}

SFRS spectra measured under resonant excitation into the peak of the exciton line at 2.947~eV look, in general, similar to the case of the non-resonant excitation, Figure~\ref{SFRS_res}(a-c). We observe also here the $1e$ and $2e$ lines with the $g$-factors $g_{\rm V}=1.78$ and $g_{\rm F}=1.66$ being close to the non-resonant values. However, their polarization properties are less prominent in both the Voigt and Faraday geometries, cf. the data collected in Table~\ref{Table-1}. 

\begin{table}[h!]
    \renewcommand{\arraystretch}{1.5}
	\centering
	
\begin{tabular}{|c|p{2.5cm}|p{3.5cm}|p{3.5cm}|}

\hline
   
    {}&Excitation&Voigt geometry&Faraday geometry\\

\hline
    
    \multirow{6}{*}{1$e$} 
    & {Non-resonant} 
    & $g_{\rm V}=1.74$ 
    \par\rule{0cm}{0.5cm} $I^{1e}_{HV}/I^{1e}_{HH}=7$ 
    & $g_{\rm F}=1.59$ 
    \par\rule{0cm}{0.5cm} $I^{1e}_{HV}/I^{1e}_{HH}=2$
    \par\rule{0cm}{0.5cm} $I^{1e}_{\sigma^+\sigma^+}/I^{1e}_{\sigma^-\sigma^+}=5$\\[2pt]
    
    \cline{2-4} 
    & {Resonant} 
    & $g_{\rm V}=1.78$ 
    \par\rule{0cm}{0.5cm} $I^{1e}_{HV}/I^{1e}_{HH}=2.5$
    & $g_{\rm F}=1.66$ 
    \par\rule{0cm}{0.5cm} $I^{1e}_{HV}/I^{1e}_{HH}=1.2$
    \par\rule{0cm}{0.5cm} $I^{1e}_{\sigma^+\sigma^+}/I^{1e}_{\sigma^-\sigma^+}=2.6$\\[2pt]
    
\hline

    \multirow{6}{*}{2$e$} 
    & {Non-resonant} 
    & $g_{\rm V}=1.74$ 
    \par\rule{0cm}{0.5cm} $I^{2e}_{HV}/I^{2e}_{HH}=0.3$ 
    & $g_{\rm F}=1.62$ 
    \par\rule{0cm}{0.5cm} $I^{2e}_{HV}/I^{2e}_{HH}=0.5$
    \par\rule{0cm}{0.5cm} $I^{2e}_{\sigma^+\sigma^+}/I^{2e}_{\sigma^-\sigma^+}=1$\\[2pt]
    
    \cline{2-4} 
    & {Resonant} 
    & $g_{\rm V}=1.76$ 
    \par\rule{0cm}{0.5cm} $I^{2e}_{HV}/I^{2e}_{HH}=0.8$ 
    & $g_{\rm F}=1.64$ 
    \par\rule{0cm}{0.5cm} $I^{2e}_{HV}/I^{2e}_{HH}=1$
    \par\rule{0cm}{0.5cm} $I^{2e}_{\sigma^+\sigma^+}/I^{2e}_{\sigma^-\sigma^+}=2.8$\\[2pt]
    
\hline

\end{tabular}
	\caption{$g$-factors of single (1$e$) and double (2$e$) electron spin-flips measured for 4ML CdSe NPLs in Voigt and Faraday geometries for resonant and non-resonant excitation. The polarization properties are specified by the ratio of the intensities measured in cross- and co-polarizations. 
	}
	\label{Table-1}
\end{table}

One can also note that under resonant excitation the ratio of intensities of the $2e$ and $1e$ lines is smaller compared to the non-resonant case, cf. Figures~\ref{SFRS_nr}(b) and \ref{SFRS_res}(a). This may be explained by a smaller probability to find doubly-charged NPLs in this case. It is a well-known property of colloidal nanocrystals synthesized by wet chemistry that the doping with donor or acceptor impurities is difficult, while nanocrystals charged by electrons and/or holes can be obtained via photocharging.\cite{Qin2012,Efros2016,Rabouw2016,Feng2017} In this case one of the carriers from the photogenerated electron-hole pair in an initially neutral nanocrystal may be captured by surface states whereas the other carrier is left inside the nanocrystal as a long-living resident carrier. For different surface state origins and experimental conditions, the photocharging can last up to one month even at room temperature.\cite{Hu2019} Similar photocharging effect was exploited in SFRS experiments on epitaxially-grown CdTe/(Cd,Mg)Te quantum wells~\cite{Debus2013} and (In,Ga)As/GaAs quantum dots.\cite{Debus2014} In these cases, additional illumination with photon energy exceeding the barrier band gap was used, which resulted in spatial separation of the photogenerated carriers between the barrier and the quantum wells/dots where the photocarriers with higher mobility were collected. This kind of illumination provides an enhancement of the $1e$ SFRS signal measured for resonant excitation, which is explained by an increase of the resident electron concentration. 

We applied weak additional illumination ($P_{\rm ill}=1.3$~Wcm$^{-2}$) to the 4ML NPLs sample. The photon energy of 3.061~eV was used together with either non-resonant or resonant excitation mentioned before, Figure~\ref{SFRS_res}(d-f). A pronounced increase of the intensities for the $1e$ and $2e$ lines by a factor of about 3 is detected for the resonant excitation. However, the illumination does not affect the SFRS intensity for the non-resonant excitation. This means that non-resonant excitation itself contributes to NPL photocharging which, however, is already saturated for the used excitation densities so that the illumination does not deliver more resident electrons. It also shows that the SFRS technique can be used as sensitive tool for detecting nanocrystal charging as well as the efficiency and dynamics of photocharging. 

The SFRS spectra of the 3ML and 5ML NPLs were measured only for resonant excitation at the exciton line peak. The 5ML sample spectrum shows both the $1e$ and $2e$ SFRS lines characterized by $g$-factor values of $g_{\rm V}=1.62$ and $g_{\rm F}=1.62$ with less strict polarization properties (Supporting Information, Figure~S4), similar to the resonant excitation of the 4ML sample. The SFRS spectrum of the 3ML sample contains only the $1e$ line, giving $g_{\rm V}=1.82$ and $g_{\rm F}=1.82$, and also demonstrates partially violated polarization selection rules for resonant excitation (Supporting Information, Figure~S5).

\begin{figure}[h!]
	\centering
	\includegraphics[width=\linewidth]{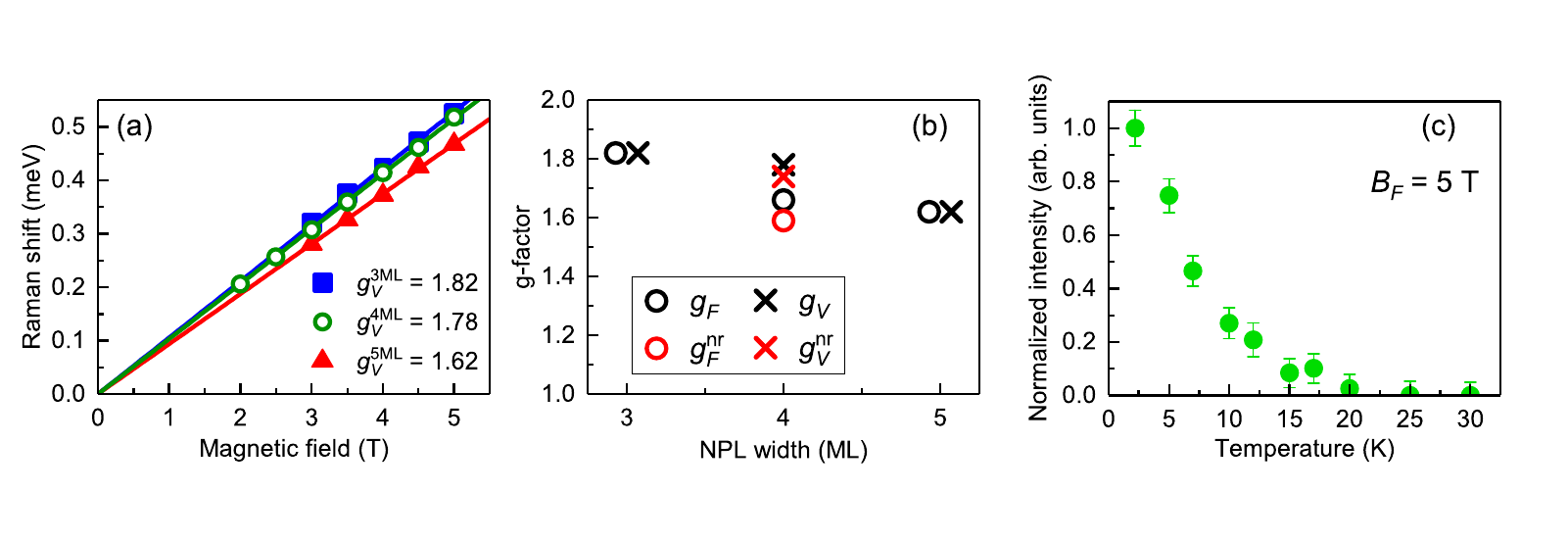}
	\caption{(a) Magnetic field dependence of the Raman shift of the single electron spin-flip line measured in the Voigt geometry for resonant excitation of the 5ML (red triangles), 4ML (green open circles), and 3ML (blue squares) samples. (b) Dependence of electron $g$-factors on the NPL thickness measured in the Voigt (crosses) and Faraday (circles) geometry for resonant (black symbols) and non-resonant (red symbols) excitation. (c) Temperature dependence of normalized integral intensity of the anti-Stokes 1$e$ line in the 4ML NPLs measured for resonant excitation in the Faraday configuration at the magnetic field of 5~T. } 
	\label{Fanplot}
\end{figure}

In order to compare the properties of NPLs of different thicknesses we plot in Figure~\ref{Fanplot}(a) the magnetic field dependencies of the $1e$ Raman shifts in the 3ML, 4ML and 5ML NPLs. The $g$-factors evaluated from the linear $B$-fits of these dependencies are collected in Figure~\ref{Fanplot}(b). Here we plot the data for both geometries and also for non-resonant excitation of the 4ML sample. The electron $g$-factor shows a noticeable but weak increase with decreasing NPL width, changing from 1.62 in the 5ML sample up to 1.82 in the 3ML NPLs.  It is known that in bulk CdSe the electron $g$-factor is positive, and the same holds for CdSe-based nanocrystals. The electron $g$-factor values measured in NPLs significantly exceed the value of 0.42 in bulk cubic CdSe~\cite{Karimov2000} due to the strong  quantum confinement of electrons which considerably increases the band gap of the NPLs. The size dependence of the electron $g$-factor was reported for CdSe spherical nanocrystals.\cite{Gupta2002,Hu2019}. In CdSe/CdS NPLs with 4ML CdSe core thickness and about 28ML thick CdS shells from each side the PL maximum at $T=2$~K is at 1.954~eV and the electron $g$-factor measured by SFRS equals to 1.68.\cite{Shornikova2018nl} In this case the $g$-factor is contributed both by the confinement effect and by the leakage of the electron wavefunction into the CdS barriers with large electron $g$ factor of $g=1.78$.\cite{Sirenko1998,Hopfield1961} In all NPL samples, the intensity of the SFRS lines is sensitive to temperature and the lines vanish for temperatures exceeding 15~K. The temperature dependence for the 4ML sample is shown in Figure~\ref{Fanplot}(c). 

To conclude this section, our experimental results evidence that the single- and double-electron SFRS lines originate from NPLs charged by resident electrons. This is supported by the presence of a trion emission line in PL, the ratio between the $1e$ and $2e$ SFRS line intensities, and the illumination effect on SFRS. As we will show below, the polarization of the SFRS lines is also in agreement with the model expectations for an exciton interacting with resident electrons.  Therefore, in the theory developed in the next Section we focus on SFRS processes in NPLs charged with resident electrons.  It is worth to stress that spin flip of an electron bound in the neutral exciton is excluded from the consideration because in the studied NPLs the electron Zeeman splitting is smaller than the bright-dark energy splitting, $\Delta E_{\mathrm{AF}}$, of the exciton states due to the electron-hole exchange interaction: according to Ref.~\cite{Shornikova2018ns}, in the 4ML CdSe NPLs $\Delta E_{\mathrm{AF}}=5$~meV, for example.
    

\section{Theory}
As mentioned in the Introduction, the experimental observation of a double-electron spin flip is very unusual. We have found only one report on its observation in SFRS spectra of bulk CdS dated back to 1972.\cite{Scott1972} The theory of multiple spin-flip Raman scattering proposed by Economou et al. and published in the same issue of the Physical Review Letters~\cite{Economou1972} is based on the interaction of two or more donor-bound electron spin states with a photoexcited exciton state. In the experiment~\cite{Scott1972} the Raman shift of the double spin-flip line was reduced by 0.25 cm$^{-1}$ relative to twice the single flip shift due to the exchange interaction of the electrons bound at a pair of neighboring donors. We have not found such an exchange contribution in our experiments on CdSe NPLs, meaning that the exchange interaction between localized resident electrons is smaller than 20~$\mu$eV.

\begin{figure}[h]
	\centering
	\includegraphics[width=15cm]{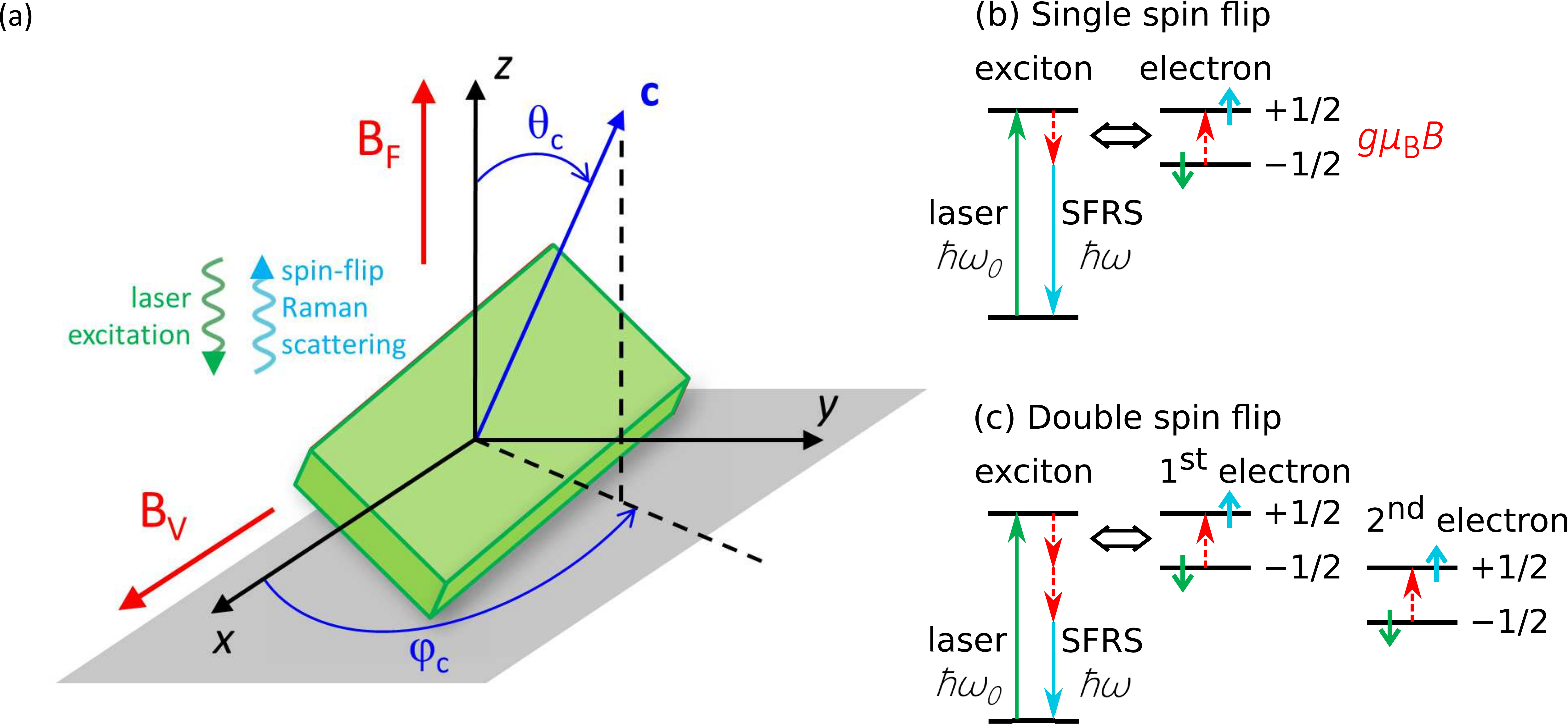}
	\caption{(a) Geometry of the experiment: The $x$, $y$ and $z$-axes represent the laboratory frame, ${\textbf{c}}$ is the unit vector normal to the NPL surface, red arrows show the directions of the external magnetic field in the Voigt and Faraday geometries. The excitation beam propagates along the $z$-axis and the SFRS signal is measured in back scattering. (b) Single electron spin-flip mechanism for the 1$e$ line and (b) double electron spin-flip mechanism for the 2$e$ line in an external magnetic field.  $\hbar \omega_0$ and  $\hbar \omega$ are the energies of incident and scattered photons, respectively.}
	\label{schema}
\end{figure}

In this section we generalize the work of Economou et al. and present a theory of the single and double spin-flip scattering processes in colloidal nanoplatelets hosting one or two localized resident electrons. As compared to Ref.~[\onlinecite{Economou1972}] we analyze the compound matrix elements of the spin-flip scattering and derive the selection rules and polarization properties of the 1$e$ and 2$e$ SFRS. The theory takes into account the anisotropy of the electron $g$ factor and an arbitrary orientation of the NPL with respect to the light propagation direction along the $z$ direction, normal to the substrate (the axes $x, y, z$ represent the laboratory frame) as shown schematically in Figure~\ref{schema}(a). We consider an arbitrary direction of the external magnetic field, the field directions for the Voigt and Faraday geometries are shown by the red arrows. The processes responsible for the SFRS signals are shown schematically in Figure~\ref{schema}(b,c) for the Stokes components.

Hereafter, we use the following notations: ${\bf c}$ for the unit vector along the normal to the NPL, $\mathbf{b}=\textbf{B}/B$ for the unit vector along the magnetic field, $g_{\parallel}$ and $g_{\perp}$ for the principal values of the electron $g$-factor tensor, longitudinal ($\textbf{B} \parallel \textbf{c}$) and transverse ($\textbf{B} \perp \textbf{c}$) to the ${\bf c}$-vector, $ \mathbf{\tilde b} = \left( g_{\perp} \textbf{b}_{\perp}  + g_{\parallel} \textbf{b}_{\parallel}\right)/g $ for the unit vector of the effective magnetic field inside the NPL, and $g$ for the effective electron Land\'{e} factor
\begin{equation} \label{effgfactor}
g= \sqrt{g_\perp^2 \sin \Theta_{B}^2+g_\parallel^2  \cos \Theta_{B}^2 }= \sqrt{g_\perp^2 +(g_\parallel^2 -g_\perp^2) ( \textbf{b} \cdot \textbf{c})^2 } \, .
\end{equation}
Thus, the Zeeman splitting of the resident electron, $g \mu_B B$, depends on the angle $ \Theta_{B}$ between the magnetic field direction and $\textbf{c}$.

In the single SFRS Stokes process (Figure~\ref{schema}(b)), the initial state comprises the incident photon of energy $\hbar \omega_0$ propagating along the $z$ axis and the resident electron in the spin-down state $\downarrow_{\mathbf{\tilde b}}$ along the direction of the effective magnetic field $\mathbf{\tilde B}=B\mathbf{\tilde b}$. At the final state, the emission of the secondary photon with energy $\hbar \omega = \hbar \omega_0 -  g \mu_{\rm B} B $ leaves the resident electron in the spin-up state. In the intermediate state, the incident photon generates an optically-allowed (bright) exciton formed by a heavy-hole with the angular momentum projection $J=\pm 3/2$ on the $c$ axis. We assume that $\Delta E_{\mathrm{AF}} \gg g \mu_B B $.
 
The intensity of the single SFRS Stokes process can be found as $I^{(1e)} \propto |V^{(1e)} |^2 \delta(\hbar \omega_0-\hbar \omega - g \mu_{\rm B} B) $.  The compound matrix element of the main process contributing to the single SFRS signal has the form
\begin{equation}  \label{ME1}
V^{(1e)} = \frac{M_{\rm em}^{(i)} M_{\rm abs} }{ E_{\rm exc} - \hbar \omega_0 - {\rm i} \hbar \Gamma}\,.
\end{equation}
Here $E_{\rm exc}$ is the energy required for resonant optical excitation of the bright heavy hole exciton. The kinetic decay parameter $\Gamma$ is the sum of the exciton damping rate and the hopping rate of the resident electron out of its localized state. The  matrix elements $M_{\rm abs}$ and  $M^{(i)}_{\rm em}$ are proportional to the interband matrix element of the momentum operator and describe, respectively,  the absorption of the incident light with generation of the exciton and the emission of the secondary photon due to radiative recombination of the resident electron $i$ with the hole in the exciton. 

It is necessary to note that for the single SFRS the presence of only one resident electron ($i=1$) is sufficient assuming that its envelope function $\phi (\mathbf{r})$  overlaps with the exciton envelope function $\Psi_{\rm exc}(\textbf{r}_e, \textbf{r}_h)$. The presence of two resident electrons ($i=1,2$) overlapping with the exciton should result in an increase of the single SFRS intensity, up to a factor of 2. In contrast, the observed double SFRS can be understood only by assuming the presence of two (or more, hereafter we assume only two) resident electrons in a NPL, see Figure~\ref{schema}(c).  The resident electrons are localized in different parts of the platelet. While their envelope functions $\phi_{1}(\mathbf{r})$ and $\phi_{2}(\mathbf{r})$ each overlap with the exciton, they overlap only weakly with each other. As a result, the singlet-triplet energy splitting of the resident electrons is smaller than the Zeeman energy and the lowest two-electron state in the external magnetic field is the triplet one $\downarrow_{1,\mathbf{\tilde b}} \downarrow_{2, \mathbf{\tilde b}}$ with parallel spins oriented along $\tilde{\bm b}$. Thus, in the double spin-flip process, the initial state comprises the incident photon and two resident electrons each with spin-down, while the final state consists of the scattered photon with energy $\hbar \omega = \hbar \omega_0 -  2 g \mu_{\rm B} B $ and two resident electrons both with spin up  $\uparrow_{1,\mathbf{\tilde b}} \uparrow_{2, \mathbf{\tilde b}}$. 

Contrary to the single SFRS, the double spin-flip process involves two intermediate states.
In the first intermediate state the photon is replaced by the exciton, and in the second intermediate state the product of the spinors in state $\downarrow_{1,\mathbf{\tilde b}} \downarrow_{2, \mathbf{\tilde b}}$ is replaced by the antiparallel spin triplet $(\downarrow_{1,\mathbf{\tilde b}} \uparrow_{2, \mathbf{\tilde b}} + \uparrow_{1,\mathbf{\tilde b}}  \downarrow_{2, \mathbf{\tilde b}})/\sqrt{2}$ due to the exchange interaction between the exciton and one of the resident electrons. The matrix element of the double SFRS defining the intensity  $I^{(2e)} \propto |V^{(2e)} |^2 \delta(\hbar \omega_0-\hbar \omega - 2 g \mu_{\rm B} B) $,  reads
\begin{equation} \label{ME2}
V^{(2e)} = \sum\limits_i \frac{M^{( \bar{i} \neq i )}_{\rm em} \Delta^{(i)}_{\uparrow \downarrow} M_{\rm abs} }{ ( E_{\rm exc} + g \mu_B B - \hbar \omega_0 - {\rm i} \hbar \Gamma') ( E_{\rm exc} - \hbar \omega_0 - {\rm i} \hbar \Gamma)}\:,
\end{equation}
where $\Gamma'$ is the damping rate of the second intermediate state.  
The matrix element $\Delta^{(i)}_{\uparrow \downarrow}$ describes the flip-stop transition between the $\downarrow_{i,\mathbf{\tilde b}} \uparrow_{\textbf{c}}$ and $\uparrow_{i,\mathbf{\tilde b}} \uparrow_{\textbf{c}}$ spin states ($i = 1$ or 2) caused by
the electron-electron exchange interaction $ {\cal \hat H}_{e\mbox{-}e} \propto \hat{\textbf{s}}_i \cdot \hat{\textbf{s}}$, where $\hat{\textbf{s}}_i$ and $\hat{\textbf{s}}$ are the spin operators for the $i$-th resident electron and the electron in the exciton (aligned along $\mathbf{c}$), respectively. Finally, $M^{(\bar{i} \neq i)}_{\rm em}$ describes the emission of the secondary photon due to radiative recombination of the second resident electron with spin $\downarrow_{\bar{i},\mathbf{\tilde b}}$ and the hole in the exciton.

We neglect in Eqs.~\eqref{ME1} and \eqref{ME2} a possible splitting of the doubly degenerate bright-exciton state of energy $E_{\rm exc}$ that might be caused by the external magnetic field or other perturbations, which is valid in particular if $g \mu_B B \ll {\rm max}\{|E_{\rm exc} - \hbar \omega_0|, \hbar \Gamma'\}$. Under this assumption we can also neglect the Zeeman energy $g \mu_B B$ in the denominator of Eq.~(\ref{ME2}) and replace $ E_{\rm exc} + g \mu_B B$ simply by $ E_{\rm exc}$. Then, after a series of transformations, we obtain the matrix elements of the single and double SFRS processes in the following general, but surprisingly simple form:
\begin{eqnarray}
   && V^{(1e)}(\textbf{e}^*,\textbf{e}^0) \propto \sin \Theta \, (\textbf{e}^*\times \textbf{e}^0)\cdot  \textbf{c}  \, , \label{Ve1}\\ \label{Ve2}
    && V^{(2e)}(\textbf{e}^*,\textbf{e}^0) \propto \sin^2 \Theta \,  (\textbf{e}^*\times \textbf{c}) \cdot (\textbf{e}^0 \times \textbf{c})=\sin^2 \Theta \, \left[ \textbf{e}^* \cdot \textbf{e}^0 - (\textbf{e}^* \cdot \textbf{c})(\textbf{e}^0 \cdot \textbf{c})  \right]  \, .  \end{eqnarray}
Here $\textbf{e}^0$ and $\textbf{e}$ are the polarization vectors of the absorbed and emitted light, respectively, and $\Theta$ is the angle between the  direction of the effective magnetic field $\tilde{\mathbf{B}}$ and $\textbf{c}$, which in turn is related to the angle  $\Theta_{B}$ between the magnetic field direction $\mathbf{B}$ and $\textbf{c}$ by
\begin{eqnarray}
   \sin \Theta = \frac{g_\perp}{g} \sin \Theta_{B} = \frac{g_\perp}{g}  \sqrt{1-(\textbf{b} \cdot \textbf{c})^2 } \, .  \nonumber
\end{eqnarray}

The matrix elements \eqref{Ve1} and \eqref{Ve2} depend on the magnetic field direction via $\sin{\Theta_{B}}$. The scalar product $\textbf{b} \cdot \textbf{c}$ that defines the angle $\Theta_{B}$ depends on the geometry of the experiment (direction of $\textbf{b}$ with respect to the light propagation direction $z$) and the orientation of the NPLs. The latter is defined by the orientation of the vector 
\begin{equation} \label{cvarphi}
\textbf{c}=(\sin \theta_c \cos \varphi_c,\,\sin \theta_c \sin \varphi_c, \, \cos \theta_c)
\end{equation}
with respect to the substrate as shown in Figure~\ref{schema}(a). The magnetic field direction for the two geometries used in the experiment is given by $\textbf{B}_{\rm F}=B \textbf{b}_{\rm F}=B(0,0,1)$ and  $\textbf{B}_{\rm V}=B \textbf{b}_{\rm V}=B(1,0,0)$, so that  $\sin \Theta_{B_{\rm F}}=-\sin \theta_c$ in the Faraday geometry and  $\sin \Theta_{B_{\rm V}}=\sqrt{1-\sin^2 \theta_c \cos^2 \varphi_c}$ in the Voigt geometry.

The vector ${\bf c}$ is also present in the scalar products in Eqs.~\eqref{Ve1} and \eqref{Ve2}, along with the polarization vectors of the absorbed and emitted photons. One can see that both the 1$e$ and 2$e$ SFRSs are forbidden for linear polarization of either the absorbed or emitted photon along the ${\bf c}$ vector. Moreover, Eq.~\eqref{Ve1} predicts strict cross-linear polarization rules for the 1$e$ SFRS and forbids its observation for NPLs with ${\bf c}$ perpendicular to the light propagation direction $z$. Simultaneously, the mutual orientation of the $\textbf{b}$ vector and the $\textbf{e}^0$ or $\textbf{e}^*$ vectors does not enter into Eqs.~\eqref{Ve1} and \eqref{Ve2}. Therefore, except for the ${\bf b}$-dependence of $\sin{\Theta}$, the model predicts isotropy of the SFRS selection rules with respect to the rotation of the magnetic field in the sample plane for the Voigt geometry.

\section{Discussion}

The theoretical predictions on the possibility of single and double SFRS signals including their intensity ratios, their polarizations in the Voigt and Faraday geometries depending on the NPL orientation and the corresponding $g$ values are summarized in Table~\ref{Table-2}. For the horizontal NPLs lying flat on the substrate, $\sin \theta_c=0$ and both the 1$e$ and 2$e$ SFRSs are forbidden in the Faraday geometry because $\sin \Theta_{B_{\rm F}}=0$ and, correspondingly, $V^{(1e)}=V^{(2e)}=0$, see Eqs.~\eqref{Ve1} and \eqref{Ve2}. However, they can be observed in the Voigt geometry, namely in cross polarization ($HV$) for the 1$e$ SFRS and in parallel polarizations ($HH$) for the 2$e$ SFRS. Note that for the SFRS signals the $HV$ and $VH$ configurations are equivalent to each other as are the configurations $HH$ and $VV$.

\begin{table}[h!]
	\centering
	\renewcommand{\arraystretch}{1.5}
	
	\begin{tabular}{|c|p{3cm}|p{3.5cm}|p{3.5cm}|}

\hline
   
    (a)&{}&Voigt geometry&Faraday geometry\\

\hline
    
    \multirow{7}{*}{1$e$} & {$\sin \theta_c =0$} \center\includegraphics[width=3cm]{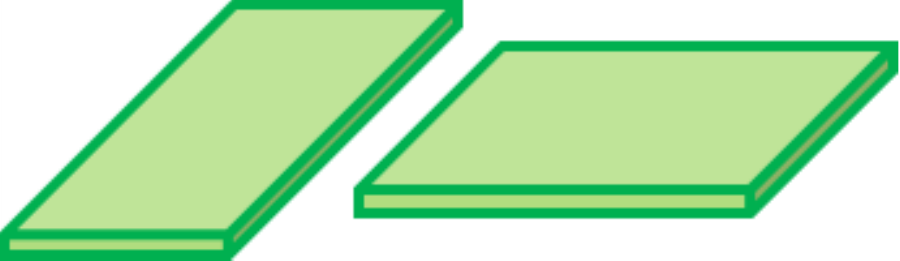} 
    & Yes 
    \par\rule{0cm}{0.5cm} $g_{\rm V}=g_\perp$ 
    \par\rule{0cm}{0.5cm} $I^{1e}_{HV}$ 
    & {No} \\[50pt]
    
    \cline{2-4} & {$\sin \theta_c =1$}
    \center\includegraphics[height=1.5cm]{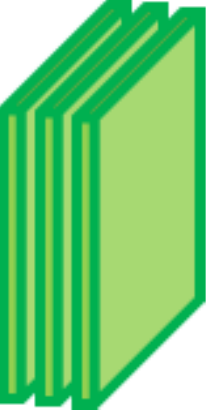} 
    & {No} 
    & {No}\\[50pt]
    
\hline

    \multirow{7}{*}{2$e$} &{$\sin \theta_c = 0$} \center\includegraphics[width=3cm]{geom_sin0.pdf} 
    & {Yes}
    \par\rule{0cm}{0.5cm} $g_{\rm V}=g_\perp$
    \par\rule{0cm}{0.5cm} $I^{2e}_{HH}$ 
    & {No} \\[50pt]
    
    \cline{2-4} & {$\sin \theta_c = 1$} \center\includegraphics[height=1.5cm]{geom_sin1.pdf} 
    & {Yes}
    \par\rule{0cm}{0.5cm} $g_\parallel < g_{\rm V} < g_\perp$ 
    \par\rule{0cm}{0.5cm} $I^{2e}_{HV}/I^{2e}_{HH}=1/7$ 
    & {Yes} 
    \par\rule{0cm}{0.5cm} $g_{\rm F}=g_\perp$ 
    \par\rule{0cm}{0.5cm} $I^{2e}_{HV}/I^{2e}_{HH}=1/3$ 
    \par\rule{0cm}{0.5cm} $I^{2e}_{\sigma^+\sigma^+}/I^{2e}_{\sigma^-\sigma^+}=1$ \\
    
\hline

\end{tabular}

\rule{0cm}{0.3cm}

\begin{tabular}{|c|p{3cm}|p{3.5cm}|}

\hline

    (b)&{}&Faraday geometry\\
    
\hline
    
    \multirow{7}{*}{1$e$} & {$\sin \theta_c \ll 1$} \center\includegraphics[width=3cm]{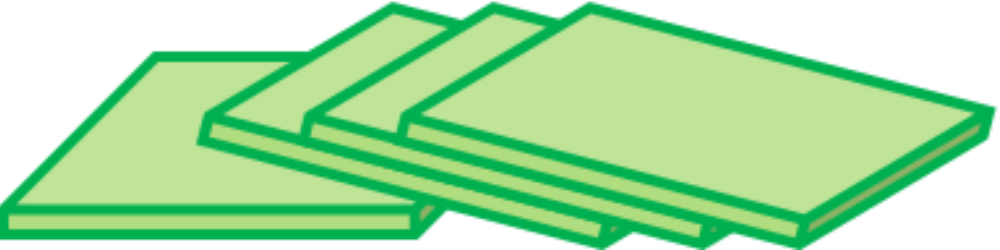} 
    & {Yes}
    \par\rule{0cm}{0.5cm} $g_{\rm F}\approx g_\parallel$ 
    \par\rule{0cm}{0.5cm} $I^{1e}_{HV}$ 
    \par\rule{0cm}{0.5cm} $I^{1e}_{\sigma^+\sigma^+}$ \\
    
    \cline{2-3} & {$\sin \theta_c \approx 1$}
    \center\includegraphics[height=1.5cm]{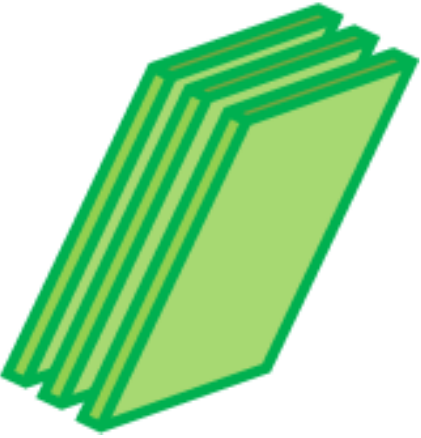} 
    & {Yes} 
    \par\rule{0cm}{0.5cm} $g_{\rm F}\approx g_\perp$
    \par\rule{0cm}{0.5cm} $I^{1e}_{HV}$ 
    \par\rule{0cm}{0.5cm} $I^{1e}_{\sigma^+\sigma^+}$ \\
    
\hline

\end{tabular}
	\caption{Theoretical predictions for the 1$e$ and 2$e$ SFRS signals: (a) for NPLs laying flat and standing straight on the substrate in the Voigt and Faraday geometries. The polarization properties are given either by the strict polarization rules or by the ratio of the intensities in the $HH$ and $HV$ configurations for linearly polarized light and in the $\sigma^+ \sigma^+$, $\sigma^+ \sigma^-$ configurations for circularly polarized light. (b) The same for tilted NPLs in the Faraday geometry.}
	\label{Table-2}
\end{table}

For the vertically oriented NPLs ($\sin \theta_c=1$) standing on their edge on the substrate, e.g., in stacks, the 1$e$ SFRS is forbidden for any direction of the magnetic field. The reason is that for light propagating along the $z$ direction, i.e., perpendicular to $\textbf{c}$, the scalar-vector product $ (\textbf{e}^*\times \textbf{e}^0)\cdot \textbf{c}$ vanishes. On the other hand, the 2$e$ SFRS can be observed for the vertical NPLs both in the Faraday and the Voigt geometry and is polarization dependent. The polarization selection rules presented in Table~\ref{Table-2} are obtained by averaging the SFRS intensity over the randomly distributed angle $\varphi_c$ in the NPL ensemble. The strongest 2$e$ SFRS signal is expected for parallel linear polarizations $HH$, while in the Faraday geometry it has the same intensity for the co- ($\sigma^+ \sigma^+$) and cross- ($\sigma^+ \sigma^-$) circular polarizations.

Let us compare the theoretical predictions from Table~\ref{Table-2} with the experimental data of Table~\ref{Table-1}(a). We notice that the observed 1$e$ and 2$e$ SFRS in the Voigt geometry can originate from horizontally lying NPLs having $g_{\rm V}=g_\perp$. Additionally, 2$e$ SFRS in the Voigt geometry can originate from vertical NPLs with the $g$-factor depending on the angle $\varphi_c$ which describes the NPL orientation with respect to the magnetic field direction, Eqs.~(\ref{effgfactor}) and (\ref{cvarphi}). According to Eq.~(\ref{effgfactor}), the effective $g$ factor in the Voigt geometry satisfies the inequalities $0 < g_\parallel \leq g_{\rm V} \leq g_\perp$, where we take into account that in the NPLs, similarly to quantum wells,  $g_\parallel<g_\perp$.\cite{Ivchenko1992} 
As the same values of $g_{\rm V}$ are observed in the 1$e$ and 2$e$ SFRS signals, Table~\ref{Table-1}, we conclude that both the 1$e$ and 2$e$ SFRS  in the Voigt geometry are mostly contributed by the horizontal NPLs.

According to Table~\ref{Table-2}(a), the 1$e$ SFRS is forbidden in the Faraday geometry for both horizontal and vertical NPLs. Nevertheless, the signal is clearly observed in experiment. This fact can be understood by analyzing the 1$e$ SFRS signal in the Faraday geometry for tilted NPLs. Table~\ref{Table-2}(b) shows theoretical predictions for two cases: NPLs slightly tilted from the horizontal geometry ($\sin \theta_c \ll 1$) and  NPLs slightly tilted from the vertical geometry ($\sin \theta_c \approx 1$). As one can see from Table~1, the experimental value of $g_{\rm F}$ is smaller than $g_{\rm V}$. Therefore, we conclude that {$g_{\rm F}  \approx g_\parallel$} and the 1$e$ SFRS in the Faraday geometry arises from NPLs slightly off-plane. In addition, in this geometry the 2$e$ SFRS with $g_{\rm F} \approx g_\parallel$ is allowed as well.  It is important to note, that even for the titled NPLs the model predicts rather strict rules for the cross linear polarizations in the 1$e$ SFRS using the Faraday geometry. 

The SFRS in the Faraday geometry arises thus from the NPLs with slightly varying small angles $\theta_c$ while the SFRS in the Voigt geometry arises from the horizontal NPLs with $\theta_c=0$. The $\theta_c$ angle dispersion in the Faraday geometry explains the larger linewidth $\delta$ of the SFRS lines as compared to the Voigt geometry, see Figure~\ref{SFRS_nr}(b,c). 

A further prediction of the theory is that the 1$e$ and 2$e$ SFRS should have opposite linear polarizations in the Voigt and Faraday geometries: the 1$e$ line is allowed only in $HV$ polarization, whereas the 2$e$ line should be strictly $HH$ polarized for the horizontal NPLs and preferably $HH$ polarized for the vertical NPLs. Also, the 1$e$ SFRS line should be strictly co-polarized for circular polarized light in the Faraday geometry. This is indeed what is observed in the experiment, more strictly for the non-resonant excitation and partly violated for the resonant excitation. Here we use the terms ``resonant'' and ``non-resonant'' for the excitation at the peak and high energy tail of the heavy-hole exciton PL spectrum, respectively. The latter excitation regime can involve the virtually excited exciton ground states or the de facto excited higher exciton states and exciton ground states within the fading inhomogeneous distribution of the NPL ensemble. Note that the non-resonant excitation used in the experiment corresponds to the photon energy $\hbar \omega_0$ about 50~meV above the PL peak, but still $\sim$100~meV below the light-hole exciton. 

The violation of the  $HV$ polarization selection rule observed for the 1$e$ SFRS, especially for resonant excitation in the Faraday geometry (Table~\ref{Table-1}), can be explained by allowing for perturbation such as distortions which can split the doubly degenerated heavy-hole exciton energy  $E_{\rm exc}$ by a small difference $\Delta$.  The sublevel splitting can be provided particularly by (i) the Zeeman splitting between the circularly-polarized heavy-hole bright exciton sublevels $\propto B \cos \Theta_B$; (ii) an in-plane anisotropy of the NPLs; and (iii) the exchange interaction between the electron in the exciton and the resident electron that maintains its spin. Explicitly accounting for these perturbations results in a correction $\delta V^{(1e)}$ to the matrix elements for the 1$e$ SFRS, lifting the strict prohibition of SFRS for linearly co-polarized excitation. The relative contribution of this correction to the SFRS matrix element can be roughly estimated as $\delta V^{(1e)}/V^{(1e)} \approx \Delta/\hbar \Gamma$. Obviously, the value of the exciton damping rate $\Gamma$ is expected to be larger for non-resonant excitation, thus providing more strict selection rules, in agreement with the experimental data. The observed violation of the $HV$ polarization selection rule for the 1$e$ SFRS in the Faraday geometry ($I_{HV}/I_{HH} =1.2$ for resonant excitation) is more pronounced than in the Voigt geometry (Table~\ref{Table-1}). This allows us to select among the above three possibilities the Zeeman splitting of the bright exciton state, see case (i), which vanishes for the horizontal NPLs in the Voigt geometry. 

Introducing the numbers $N_1$ and $N_2$ of the NPLs with one or two resident electrons we can roughly estimate the ratio of intensities of the 1$e$ and 2$e$ SFRS lines by  
$$ \frac{N_2}{{\rm max}\{N_1,N_2\}} \left\vert\frac{V^{(2e)}}{V^{(1e)}}\right\vert^2 \sim \frac{N_2}{{\rm max}\{N_1,N_2\}} \frac{\left\vert \Delta_{\uparrow \downarrow}\right\vert^2}{(E_{\rm exc} - \hbar \omega_0)^2 + (\hbar \Gamma)^2}\:,$$
where the matrix element $\Delta_{\uparrow \downarrow}$ is defined in Eq.~(\ref{ME2}). 

The dependence of the SFRS intensity on temperature shown in Figure~\ref{Fanplot}(c) can be related to the strong temperature dependence of the damping rate $\Gamma$. The increase of $\Gamma$ with increasing temperature can be caused by activation of the hopping of the resident electrons between localized sites within the NPL (or between the NPLs). 

In conclusion, we have identified, both experimentally and theoretically, that the unusual double-electron spin-flip signal is provided by nanoplatelets charged with more than one resident electron. We have also demonstrated that spin-flip Raman scattering is a powerful tool for investigating the spin level structure and spin-dependent phenomena in colloidal nanocrystals. It allows one to measure the electron $g$-factor and its anisotropy by varying the orientation of the external magnetic field and analyzing the polarization properties of the SFRS signals. The technique has been tested and approved on ensembles of CdSe nanoplatelets of different thickness. The developed theory shows that the most efficient mechanism for electron SFRS involves the interaction of a photogenerated exciton with one or two resident electrons. Therefore, the technique can serve as  valuable tool to study charging and photocharging of colloidal nanocrystals. Spin-flip Raman scattering can be also used for studying other types of colloidal nanocrystals, like quantum dots or quantum rods from different materials.

\section{Methods}

\textbf{Samples.}
The CdSe NPLs were synthesized according to the protocol reported in Ref.
~[\onlinecite{Ithurria2008}] in argon atmosphere. They have zinc-blende crystalline structure. Three batches of CdSe NPLs with thicknesses of 3, 4 and 5 monolayers (ML) were studied. Transmission electron microscopy images of the 4ML and 5ML samples are shown in the Supporting Information, see Figure~S1. Samples for optical experiments were prepared by drop-casting of a concentrated NPL solution onto a Silicon substrate. The resulting NPL ensembles contain NPLs with various orientations.

\textbf{Spin-flip Raman scattering (SFRS).}
For excitation of the SFRS we used emission lines of an Ar-ion (486.5~nm, 488~nm, 514.5~nm), a He-Cd (441.6~nm) or a Nd:YAG (532~nm) lasers. The laser power density on the sample surface did not exceed 26~Wcm$^{-2}$. The scattered light was analyzed by a Jobin-Yvon U1000 double monochromator equipped with a cooled GaAs photomultiplier connected to conventional photon counting electronics. To record sufficiently strong SFRS signal and to suppress the laser stray light, spectral slit widths of 0.2~cm$^{-1}$ (0.024~meV) were used. Most of the measurements were performed on samples immersed in pumped liquid helium (typically at a temperature of 2~K), while the measurements of the temperature dependence of the SFRS intensity were done in helium exchange gas. A split-coil superconducting solenoid was used to generate magnetic fields, $B$, up to 5~T. The SFRS was measured in backscattering geometry. The magnetic field was applied either parallel to the light wave vector (Faraday geometry) or perpendicular to it (Voigt geometry). To characterize the polarization properties of the SFRS lines in Faraday configuration ($\textbf{B}\parallel z$) we use the notation $z(\sigma^\eta,\sigma^\lambda)\bar{z}$ with $\bar{z}$ and $z$ indicating orientations perpendicular to the sample plane $xy$ and $\eta = \pm$, and $\lambda = \pm$ denoting the circular polarization of the exciting $\sigma^\eta$ and the scattered $\sigma^\lambda$ light. Here the signs $\eta$ and $\lambda$ are determined by the sign of the projection of the angular momentum of the photons on the propagation direction of the exciting light ($z$ direction). To describe the polarization properties of the SFRS spectra measured in Voigt geometry we use the notations $H$ and $V$, which correspond to the photon electrical vector $\bf E$ parallel (horizontal $H$) and perpendicular (vertical $V$) to the magnetic field direction, respectively.

\textbf{ASSOCIATED CONTENT}

\textbf{Supporting Information.}
Transmission electron microscopy (TEM) images of NPLs, photoluminescence and SFRS spectra for NPLs of various thicknesses.

\textbf{AUTHOR INFORMATION}

Corresponding Authors:

dmitri.yakovlev@tu-dortmund.de, sapega.dnm@mail.ioffe.ru, anna.rodina@mail.ioffe.ru

\textbf{ORCID}


Victor F. Sapega:  0000-0003-3944-7443

Dmitri R. Yakovlev: 0000-0001-7349-2745

Ina V. Kalitukha:  0000-0003-2153-6667

Elena V. Shornikova: 0000-0002-6866-9013

Anna V. Rodina:  0000-0002-8581-8340


Michel Nasilowski:  0000-0002-2484-7674

Benoit Dubertret:   0000-0002-9450-8029

Manfred Bayer:  0000-0002-0893-5949

\textbf{Notes}

The authors declare no competing financial interests.

\textbf{ACKNOWLEDGEMENTS}

The authors are thankful to L. Biadala, A. A. Golovatenko, Yu. G. Kusrayev, and Al. L. Efros for fruitful  discussions.  We acknowledge support by the Deutsche Forschungsgemeinschaft via the International Collaborative Research Centre TRR 160 (Projects B1 and B2) and the Russian Foundation for Basic Research (Project No. 19-52-12064 NNIO-a).

\clearpage
\setcounter{equation}{0}
\setcounter{figure}{0}
\setcounter{table}{0}
\setcounter{section}{0}
\setcounter{page}{1}
\renewcommand{\theequation}{S\arabic{equation}}
\renewcommand{\thefigure}{S\arabic{figure}}
\renewcommand{\thetable}{S\arabic{table}}
\renewcommand{\thesection}{S\arabic{section}}

\begin{center}
	\textbf{\large Supplementary Information:}
	
	\vspace{3mm}	
	\textbf{\large Single and double electron spin-flip Raman scattering in CdSe colloidal nanoplatelets}
	
	\vspace{3mm}
	
	Dennis Kudlacik,$^{1}$ Victor F. Sapega,$^{2}$ Dmitri R. Yakovlev,$^{1,2}$ Ina V. Kalitukha,$^{2}$ Elena V. Shornikova,$^{1}$ Anna V. Rodina,$^2$ Eugeniius L. Ivchenko,$^2$ Grigorii S. Dimitriev,$^2$ Michel Nasilowski,$^3$ Benoit Dubertret,$^3$  and Manfred Bayer$^{1,2}$
\end{center}
\vspace{3mm}

{\small \noindent$^1$Experimentelle Physik 2, Technische Universit{\"a}t Dortmund, 44227 Dortmund, Germany
	
	\noindent$^2$Ioffe Institute, Russian Academy of Sciences, 194021 St. Petersburg, Russia
	
	\noindent$^4$Laboratoire de Physique et d'Etude des Mat\'{e}riaux, ESPCI, CNRS, 75231 Paris, France
	
	\section{S1. TEM of CdSe nanoplatelets}
	
	\begin{figure}[h]
		\centering
		\includegraphics[width=12cm]{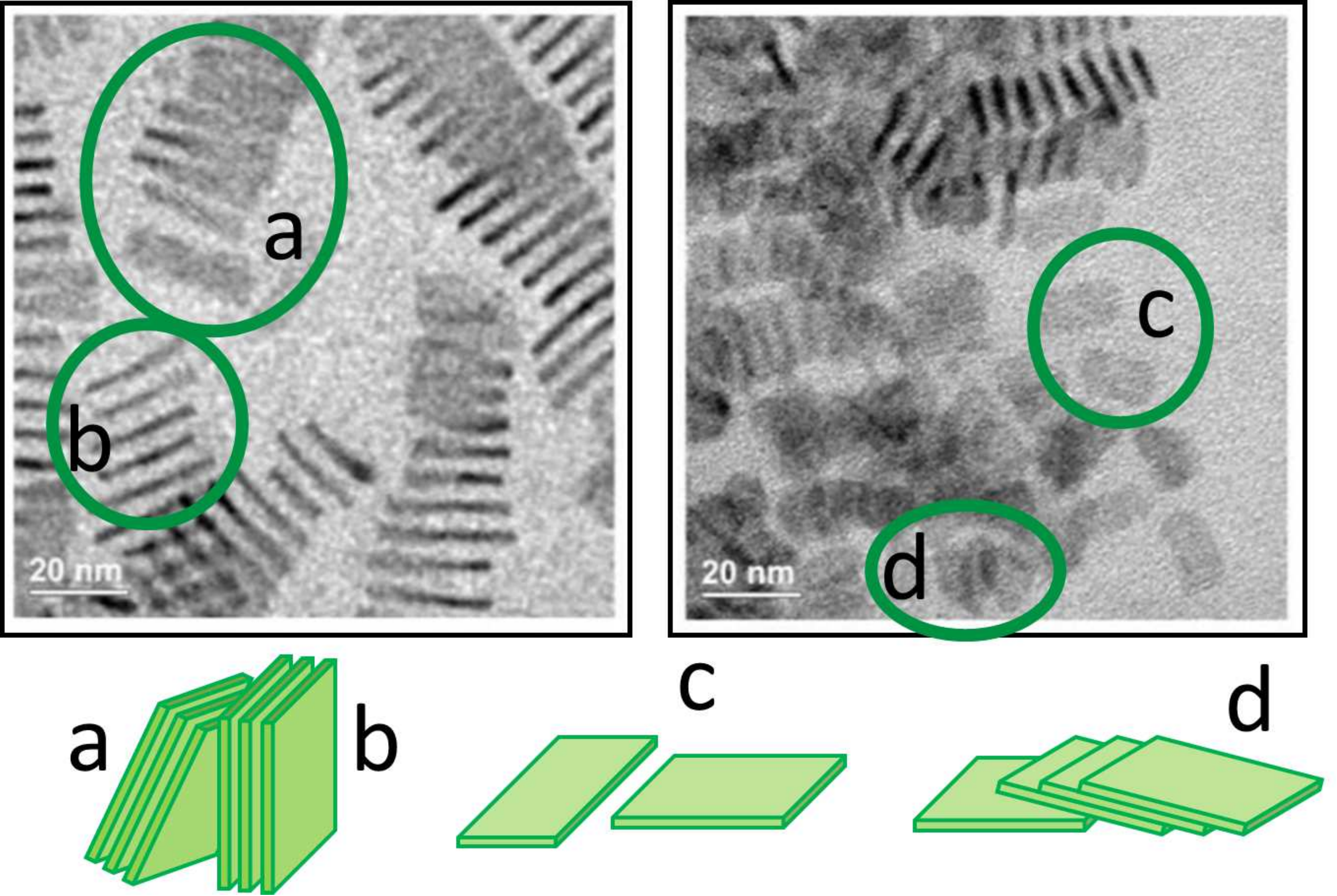}
		\caption{Transmission electron microscopy (TEM) images of the 5ML (left image) and 4ML (right image) CdSe NPL samples. Different areas on the images are marked according to the NPLs orientation on the substrate: (a) tilted stack on the substrate; (b) stack standing straight on the substrate; (c) NPLs laying flat on the substrate; and (d) slightly tilted NPLs.}
		\label{TEM}
	\end{figure}
	
	\clearpage
	
	\section{S2. Photoluminescence of CdSe nanoplatelets}
	
	Photoluminescence (PL) spectra of the 3ML, 4ML and 5ML NPLs are shown in Figure~\ref{PL}. The spectra of all samples consist of two lines, with the high-energy one (X) attributed to the neutral exciton emission and the low-energy line (T) related to recombination of the negatively charged excitons (trions). This line identification is based on the spectral shifts in emission and absorption, the characteristic recombination dynamics along with its dependence on temperature and magnetic field, and the polarization properties of emission in high magnetic fields. Details can be found in Refs.~\onlinecite{Shornikova2018ns,Shornikova2018nl,Shornikova2019nn}.
	\begin{figure}[h!]
		
		\centering
		\includegraphics[width=12cm]{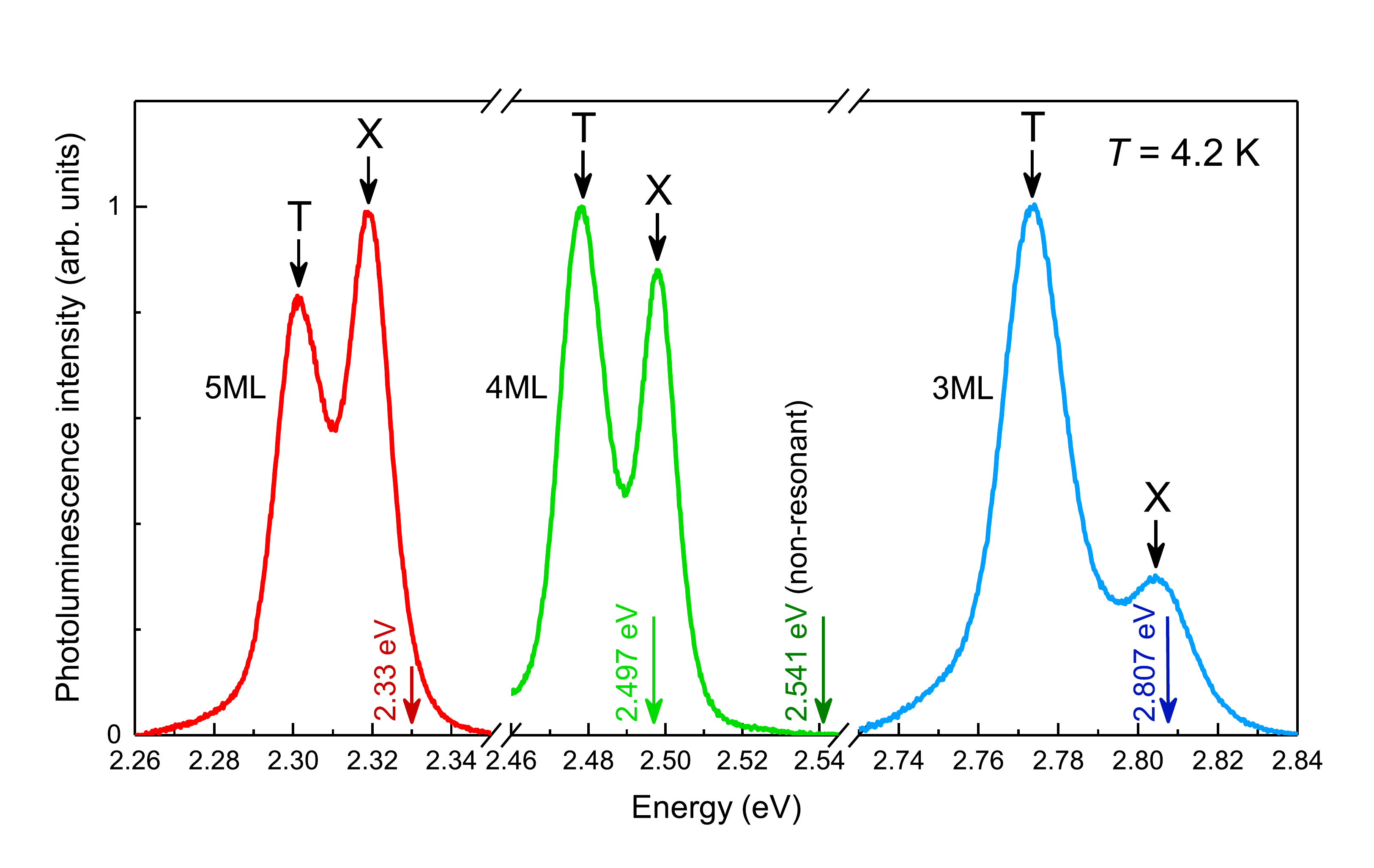}
		\caption{Photoluminescence spectra of CdSe NPLs of 3ML, 4ML and 5ML thickness measured for excitation with photon energy at 3.061~eV. $T=4.2$~K. Lines marked X and T correspond to emission of neutral excitons and negatively charged excitons (trions), respectively. Colored arrows indicate the laser energies used in the SFRS experiments.}
		\label{PL}
	\end{figure}
	
	\clearpage
	
	\section{S3. SFRS spectra of 4ML CdSe nanoplatelets}
	\begin{figure}[h]
		\centering
		\includegraphics[width=10cm]{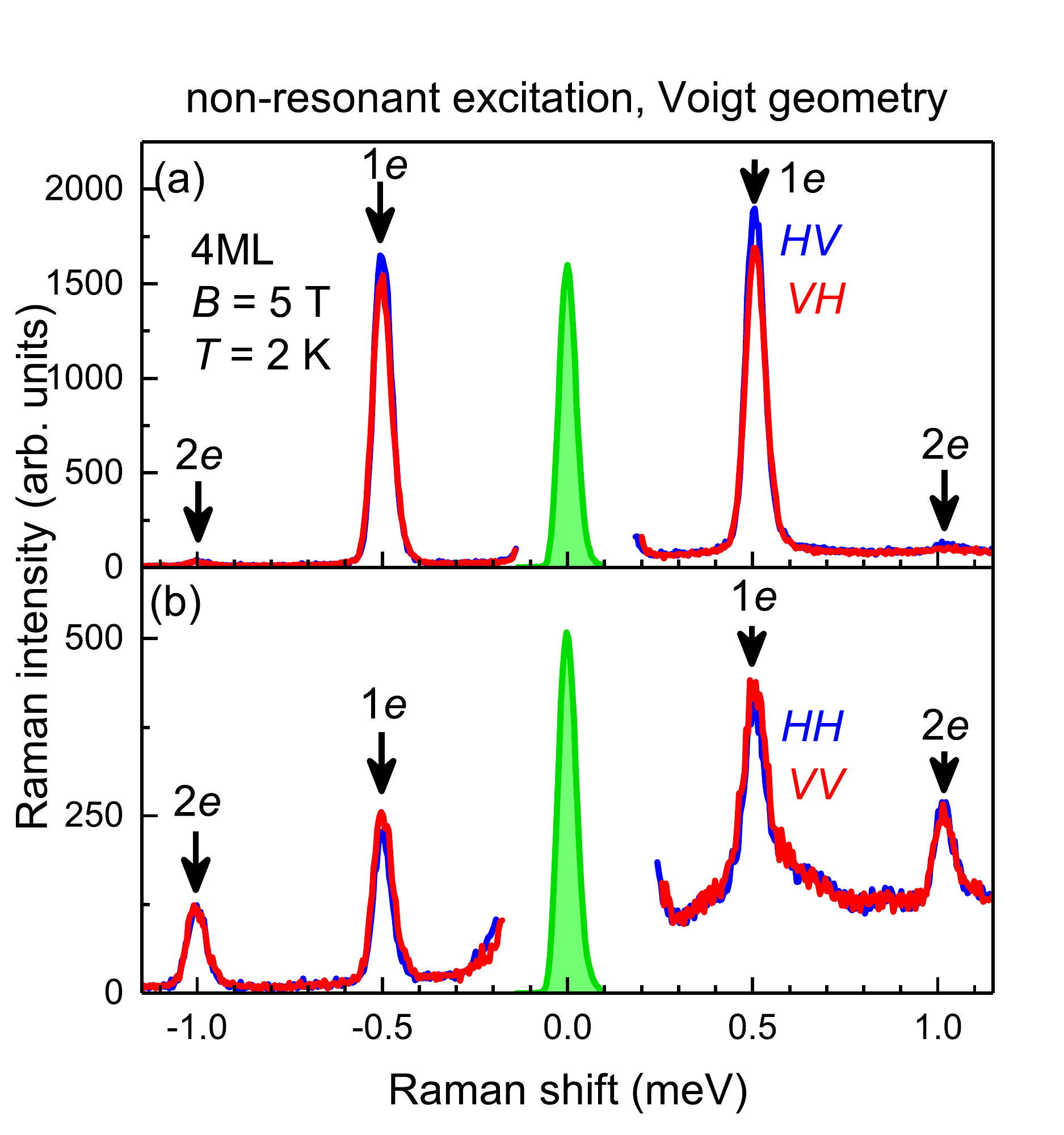}
		\caption{SFRS spectra of the 4ML CdSe NPLs measured for non-resonant excitation at 2.541~eV in a magnetic field of $B=5$~T applied in Voigt geometry at $T=2$~K. One can see in panel (b) that the co-polarized configurations $HH$ and $VV$ are equivalent in our experiment. The same holds for the case of cross-polarizations $HV$ and $VH$ shown in panel (a). Green area shows the laser line.
		}
		\label{SFRS4ML}
	\end{figure}
	
	\clearpage
	
	\section{S4. SFRS spectra of 5ML CdSe nanoplatelets}
	
	\begin{figure}[h]
		\centering
		\includegraphics[width=10cm]{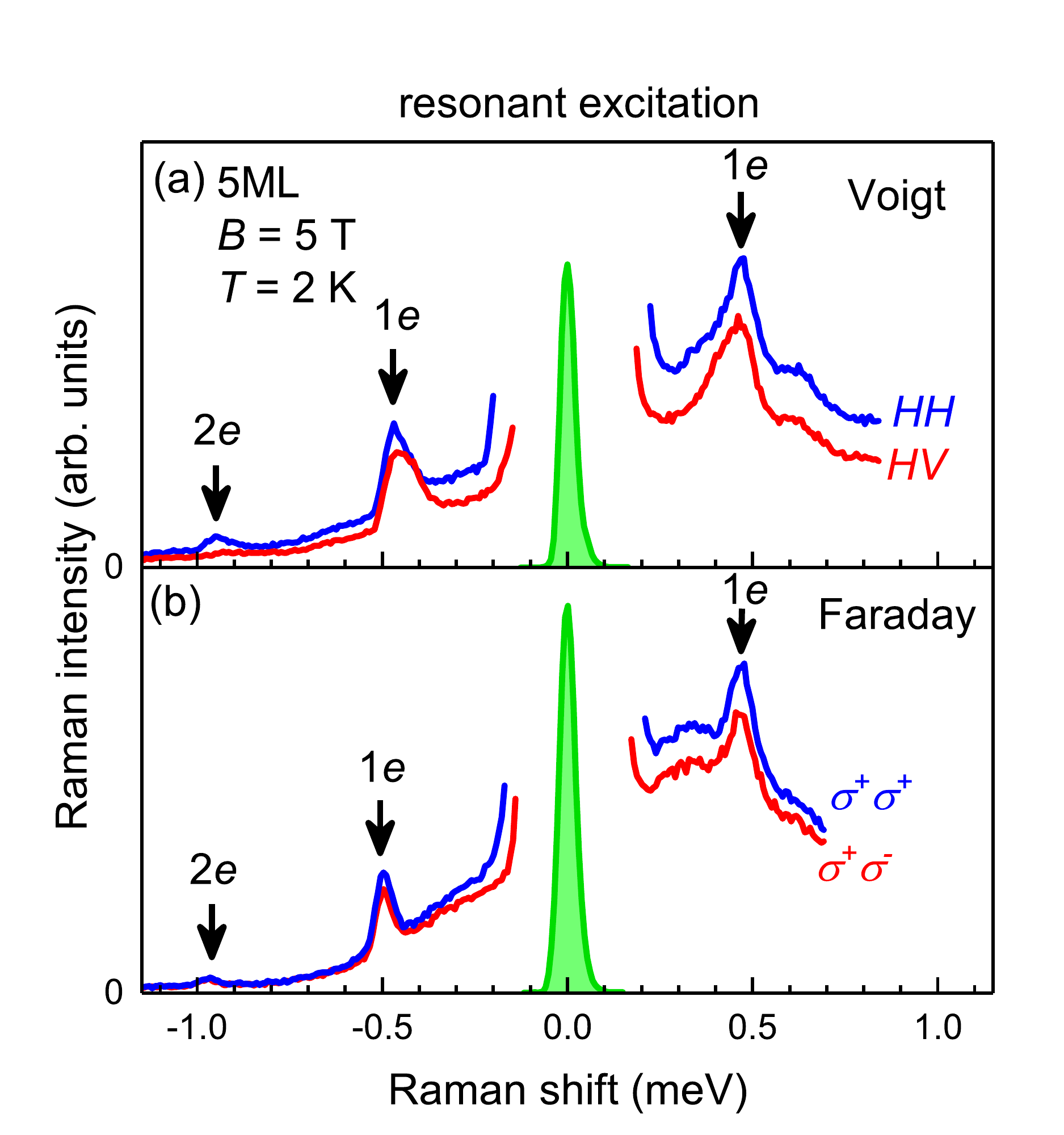}
		\caption{SFRS spectra of the 5ML CdSe NPLs measured for resonant excitation at 2.33~eV (see red arrow in Fig.~\ref{PL}) in a magnetic field of $B=5$~T applied in Voigt (a) and Faraday (b) geometries at $T=2$~K. In Voigt geometry the spectra were measured in co- (blue) and cross- (red) linear polarizations. In Faraday geometry the spectra were measured in co- (blue) and cross- (red) circular polarizations. Green line is the laser.}
		\label{SFRS5ML}
	\end{figure}
	
	\clearpage
	
	\section{S5. SFRS spectra of 3ML CdSe nanoplatelets}

	\begin{figure}[h]
		\centering
		\includegraphics[width=10cm]{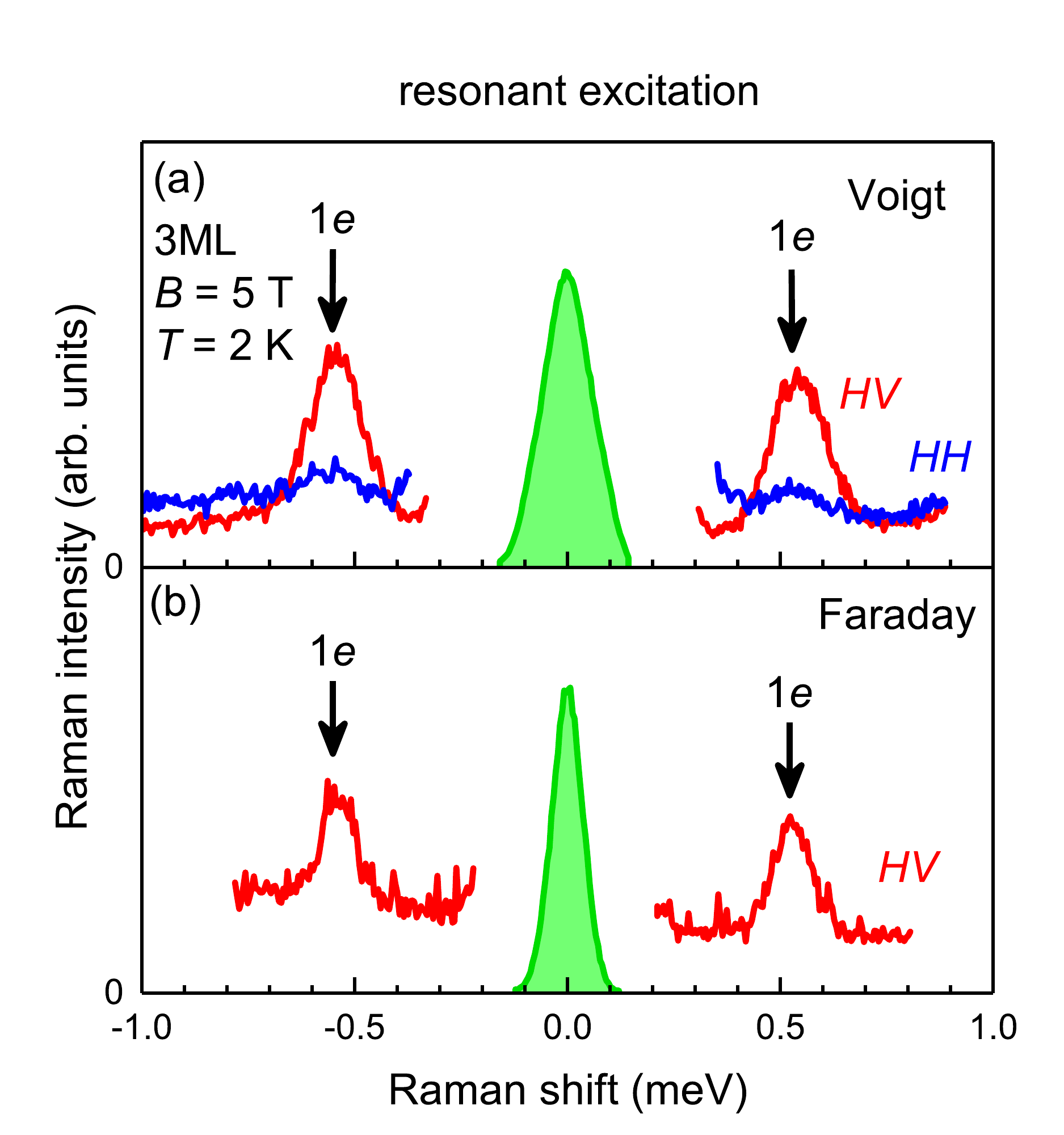}
		\caption{SFRS spectra of the 3ML CdSe NPLs measured for resonant excitation at 2.807~eV (see blue arrow in Fig.~\ref{PL}) in a magnetic field of $B=5$~T applied in Voigt (a) and Faraday (b) geometries at $T=2$~K. In Voigt geometry the spectra were measured in co- (blue) and cross- (red) linear polarizations. In Faraday geometry the spectra were measured in crossed (red) linear polarizations. Green line is the laser.
		}
		\label{SFRS3ML}
	\end{figure}

\end{document}